\documentclass[aps,pra,amsmath,amssymb,superscriptaddress,onecolumn,10pt,longbibliography]{revtex4-2}

\usepackage{bm}
\usepackage{graphicx}
\usepackage{hyperref}

\def\br{\bm{\rho}}

\def\ee{\mathrm{e}}
\def\dd{\mathrm{d}}
\def\ii{\mathrm{i}}

\begin{document}

\title{3D correlation imaging for localized phase disturbance mitigation}

\author{Francesco V. Pepe}
\author{Milena D'Angelo}

\affiliation{Dipartimento di Fisica, Universit\`{a} di Bari, I-70126 Bari, Italy}
\affiliation{INFN, Sezione di Bari, I-70125 Bari, Italy}

\begin{abstract}
Correlation plenoptic imaging is a procedure to perform light-field imaging without spatial resolution loss, by measuring second-order spatio-temporal correlations of light. We investigate the possibility to use correlation plenoptic imaging to mitigate the effect of a phase disturbance in the propagation from the object to the main lens. We assume that this detrimental effect, that can be due to a turbulent medium, is localized at a specific distance from the lens, and is slowly varying in time. The mitigation of turbulence effects has already fostered the development of both light-field imaging and correlation imaging procedures. Here, we aim at merging these aspects, proposing a correlation light-field imaging method to overcome the effects of slowly varying turbulence, without the loss of lateral resolution, typical of traditional plenoptic imaging devices.
\end{abstract}


\maketitle

\section{Introduction}

The class of plenoptic (or light-field) imaging methods includes protocols and devices that are aimed to detect the \textit{light field}, namely the joint information, within the limits of wave optics, on the light spatial distribution and propagation direction \cite{adelson1992single,ng2005light,ng2005fourier,wu2017light,lam2015computational}. Interestingly, plenoptic imaging allows retrieving such a three-dimensional information retrieval in a single intensity acquisition, while making use of no interferometric measurements. This technique is currently used in diverse scientific and technical application fields, which include microscopy \cite{microscopy1,microscopy2,microscopy3,microscopy4}, stereoscopy \cite{adelson,muenzel,levoy}, wavefront sensing \cite{thesis_wu,eye,atmosphere1,atmosphere2,ko2017comparison}, particle tracking and sizing \cite{tracking}, particle image velocimetry \cite{piv}, and 3D neuronal activity functional imaging \cite{microscopy4,skocek2018highspeed}. Through the acquired directional information, plenoptic devices open the possibility to perform viewpoint changes, refocusing at different axial distances and, consequently, three dimensional reconstruction of the a finite volume \cite{3dimaging}. In state-of-art-devices, the key to achieve plenoptic imaging is to combine the main lens with an array of micro-lenses \cite{ng2005light,georgiev2010focused}, which encode on the detector a composite information on the light-field. However, in all the implementations based on first-order intensity measurement, the drawback of such a structure is a limitation of the lateral resolution, that becomes inversely proportional to the gain in directional resolution, making the Rayleigh limit set by the numerical aperture of the main lens unreachable. 

In the context of intensity correlation imaging \cite{pittman1995optical,bennink2002two,valencia2005two,gatti2004ghost,scarcelli2006can,osullivan2010comparison,brida2011systematic,cassano2017spatial,dangelo2017characterization}, a way to overcome this practical limitation emerged in a technique named correlation plenoptic imaging (CPI), capable of performing plenoptic imaging without spatial resolution loss by measuring second-order spatio-temporal correlations of light \cite{cpi_prl,cpi_exp}. Thanks to the correlation between light beams of either chaotic light or entangled photons \cite{cpi_prl,cpi_qmqm,cpi_technologies,cpi_jopt,cpiap,scagliola2020correlation,massaro2022effect,massaro2022refocusing,massaro2023correlated}, such a measurement encodes information not only on the spatial distribution of light in a given plane in the scene, but also on the direction of light. This method entails a relevant qualitative and quantitative mitigation of the trade-off between spatial and directional resolution, opening the way to a dramatic performance increase in terms of volumetric resolution, which is especially relevant when correlation measurements are performed through high-speed spatially resolving detectors \cite{massaro2023correlated,abbattista2021towards}.

In this article, we investigate the possibility to exploit CPI to mitigate the effect of a phase disturbance in the propagation from the object to the main lens. We assume that this detrimental effect, that can be due to a turbulent medium, is localized at a specific distance from the lens, and is slowly varying in time. The presence of turbulence is an outstanding challenge of imaging, that fosters the research for new tools and devices \cite{michael1996imaging}. Interestingly, one of these attempts at mitigating turbulence involves standard light-field imaging, in view of its possibility to detect the direction of light coming from specific points of the scene \cite{wu2015imaging,wu2016imaging}. On the other hand, much research has been devoted to determining the robustness to turbulence of correlation imaging methods \cite{cheng2009ghost,li2010ghost,chan2011theoretical,hardy2011reflective,dixon2011quantum,meyers2011turbulence,shi2012adaptive}. Our aim here is to merge these aspect, proposing a correlation light-field imaging method to overcome the effects of slowly varying turbulence, without the loss of lateral resolution, typical of traditional plenoptic imaging devices. 

The article is organized as follows. In Section~2, we show the detrimental effect of an axially localized phase disturbance on the first-order image collected by a standard imaging device. In Section~3, we discuss how to exploit correlation imaging to obtain a collection of sharp images of an object by CPI, despite the presence of the considered turbulence. In Section~4, we comment on how to maximize the information extraction from the CPI measurements, and how the technique can be integrated with innovative sensors and data analysis protocols.

\section{First-order imaging with an axially localized turbulence}

\begin{figure}
    \centering
    \includegraphics[width=0.8\textwidth]{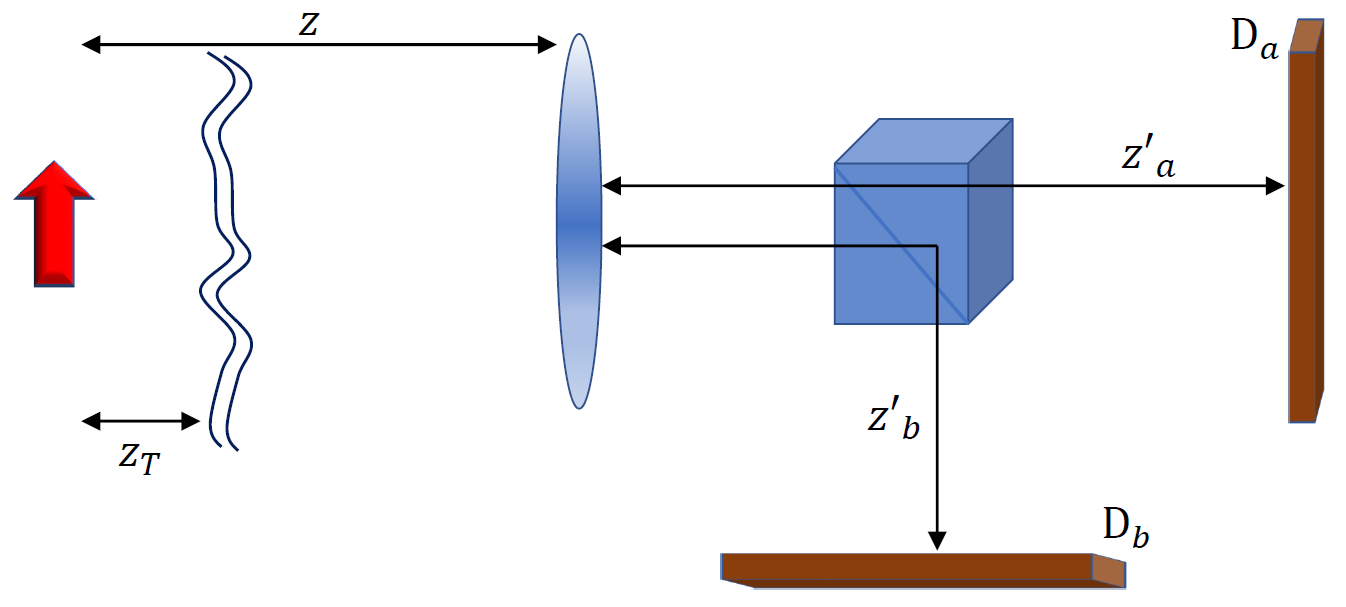}
    \caption{Optical scheme of the setup described in the article. Light from an object, treated as a chaotic emitter, encounters turbulence in a thin region of space, at a distance $z_T$, before impinging on the lens. After the lens, light is split in two optical paths, impinging on the spatially-resolving sensors $\mathrm{D}_a$ and $\mathrm{D}_b$, respectively. The lens-to-sensor distances $z'_j$, with $j=a,b$, define two object planes at the respective conjugate distances $z_j$, satisfying $1/z_j+1/z'_j=1/f$, with $f$ the focal length of the lens. The signal is collected by measuring pixel-by-pixel correlations between intensity fluctuations on the two sensors.}
    \label{fig:scheme}
\end{figure}

The optical scheme that we considered in this article, shown in Fig.~\ref{fig:scheme}, is based on the paradigm of correlation plenoptic imaging between arbitrary planes (CPI-AP) \cite{cpiap}. It consists in a single lens of focal length $f$, that collects chaotic light emitted by the object, placed at an arbitrary distance $z$, which undergoes turbulence at a distance $z_T$ on its way to the lens. After the lens, light is split in two paths by a beam splitter. The two beams impinge on the sensors $\mathrm{D}_a$ and $\mathrm{D}_b$, placed at distances $z'_a$ and $z'_b$, respectively. These two distances define two object planes at distances $z_a$ and $z_b$ in front of the lens, through the equations
\begin{equation}
\frac{1}{z_j} + \frac{1}{z'_j} = \frac{1}{f}, \quad \text{with } j=a,b.
\end{equation}
The results here presented are based on the following assumptions on turbulence:
\begin{itemize}
    \item Turbulence occurs upstream of the beam splitter, i.e.\ in the region where the optical paths are superposed.
    \item Turbulence is slowly varying in time, and can be approximated as quasi-static in the time required to perform a reasonable reconstruction of the correlation function.
    \item Turbulence is due to unpredictable perturbations of the refractive index in a region of space of small longitudinal extension $\Delta z$, such that its effect amounts to multiplying the field in that region by a phase factor $\exp(\ii k \Delta n(\br_T)\Delta z)$, with $\br_T$ the transverse coordinate of the turbulence plane and $\Delta n(\br_T)$ the change in refractive index with respect to the background. Notice, instead, that turbulence in a thick region should be described by a convolution of the field \cite{fante1985wave}.
\end{itemize}

To explain the detrimental effect of turbulence, it is instructive to start from the first-order results, namely the intensity measured at the end of one of the optical paths. Neglecting, for the sake of simplicity, the finite aperture of the lens, the field propagator from an object point of coordinate $\br_o$ and a point of coordinate $\br_a$ on the sensor $\mathrm{D}_j$ reads
\begin{multline}\label{gj}
g_j(\br_j,\br_o) = C_j \ee^{\ii k \psi(\br_j)} \exp\left( \frac{\ii k \br_o^2}{2 z_T} \right) \\ \times \int \dd^2\br_T \exp\left\{ \ii k \left[ \left( \frac{1}{z_T} - \frac{1}{z_T + z_j - z} \right) \frac{\br_T^2}{2} - \frac{\br_T}{z_T} \cdot \left( \br_o + \frac{z_T}{z_T+z_j-z} \frac{\br_j}{M_j} \right) - \phi(\br_T) \right] \right\} ,
\end{multline}
where $\phi(\br_T)=\Delta n(\br_T)\Delta z$ determines the phase change due to turbulence, and $M_j=z'_j/z_j$, with $j=a,b$, are the absolute magnifications of the images focused by the lens on the two sensors. If the object is a chaotic-light emitter with intensity profile $\mathcal{A}(\br_o)$, the intensity measured on each sensor $\mathrm{D}_j$ reads
\begin{equation}\label{firstorder}
I_j (\br_j) = \int \dd^2 \br_o \mathcal{A}(\br_o) \left| g_j(\br_j,\br_o)  \right|^2 .
\end{equation}
In the case in which we are most interested, when the object is focused on $\mathrm{D}_j$ (namely, $z=z_j$), the intensity reads, up to irrelevant constants,
\begin{equation}
I_j (\br_j) = \int \dd^2 \br_o \mathcal{A}(\br_o) \left| \int \dd^2\br_T \exp\left\{ \ii k \left[ \phi(\br_T) - \frac{1}{z_T}\br_T \cdot \left( \br_o + \frac{\br_j}{M_j} \right)    \right] \right\} \right|^2,
\end{equation}
leading to the stigmatic image $\mathcal{A}(-\br_j/M_j)$, inverted and magnified by $M_j$, in the case of no turbulence. When turbulence is present, we can characterize its effect by considering the limit of geometrical optics $k\to\infty$, and the stationary-phase approximation \cite{saleh2007fundamentals}, leading to
\begin{equation}\label{firstorder_geo}
I_j (\br_j) \sim \int \dd^2 \br_T \mathcal{A}\left( - \frac{\br_j}{M_j} + z_T \bm{\nabla}\phi(\br_T) \right) .
\end{equation}
Therefore, in the presence of turbulence, rays passing from a point $\rho_T$ on the turbulence plane are deviated by an amount proportional to the gradient of the phase disturbance $\phi$, leading to the superposition of sub-images characterized by different shifts. This situation closely resembles the case of an out-of-focus object, which is commonly tackled by plenoptic imaging, with the relevant difference that, in the case of turbulence, the image shift is unpredictable a priori. Notice that the effect of turbulence is less and less relevant as it occurs closer to the sample ($z_T\to 0$), as the spread due to turbulence at some point becomes less important that the natural point-spread due to the finite lens aperture.

\section{CPI for turbulence mitigation}

Intuitively, as it occurs in CPI to reconstruct out-of-focus images, if correlation measurement is able to detect each single sub-image with a well-defined shift, the detrimental effect of the superposition of shifted images in Eq.~\eqref{firstorder_geo} can be avoided. Therefore, let us consider the correlation function $\Gamma(\br_a,\br_b)$ between intensity fluctuations in the couple of points $\br_a$ on the sensor $\mathrm{D}_a$ and $\br_b$ on $\mathrm{D}_b$. Treating the object as a chaotic light emitter, as in Eq.~\eqref{firstorder}, one obtains
\begin{equation}\label{secondorder}
\Gamma (\br_a,\br_b) = \left| \int \dd^2 \br_o \mathcal{A}(\br_o)  g_a(\br_a,\br_o) g_b^*(\br_b,\br_o)   \right|^2 .
\end{equation}
Before treating the plenoptic-based reconstruction of the image, it it worth observing that the autocorrelation of intensity fluctuations
\begin{equation}
\Gamma (\br_a,\br_b) = I_a^2(\br_a)
\end{equation}
coincides with the squared first-order intensity. This result could entail an apparent mitigation of turbulence in second-order measurement. However, the average result of autocorrelation measurement is indistinguishable from measuring intensity and squaring the result; therefore, an effective mitigation occurs only if noise affecting $\Gamma(\br_a,\br_b)$ is lower than the one affecting $I_a^2(\br_a)$. This will be the object of further analysis.

Going back to the general form of $\Gamma$, the result for the setup in Fig.~\ref{fig:scheme} can be evaluated by inserting the propagators \eqref{gj}, yielding, up to irrelevant constants,
\begin{equation}
\Gamma (\br_a,\br_b) = \left| \int \dd^2 \br_o \int \dd^2 \br_T \int \dd^2 \br'_T \mathcal{A}(\br_o)  \exp\left[\ii k \Psi(\br_a,\br_b;\br_o,\br_T,\br'_T) \right]   \right|^2 ,
\end{equation}
where
\begin{multline}
\Psi(\br_a,\br_b;\br_o,\br_T,\br'_T) = \left( \frac{1}{z_T} - \frac{1}{z_T+z_a -z} \right) \frac{\br_T^2}{2} - \left( \frac{1}{z_T} - \frac{1}{z_T+z_b -z} \right) \frac{{\br'_T}^2}{2} \\ - \frac{1}{z_T}\br_o \cdot (\br_T - \br'_T) 
- \frac{1}{z_T+z_a -z} \br_T\cdot \frac{\br_a}{M_a} + \frac{1}{z_T+z_b -z} \br'_T\cdot \frac{\br_b}{M_b} + \phi(\br_T) - \phi(\br'_T).
\end{multline}
The stationary-phase conditions
\begin{equation}
    \bm{\nabla}_{\br_0}\Psi = \bm{\nabla}_{\br_T}\Psi = \bm{\nabla}_{\br'_T}\Psi = 0,  
\end{equation}
that determine the point(s) $(\bar{\br}_o,\bar{\br}_T,\bar{\br}'_T)$ providing dominant contribution in the geometrical-optics limit $k\to\infty$, do not form a linear system in the integration variables, as it would occur in absence of turbulence, due to the presence of $\bm{\nabla}\phi$. However, the system is solvable anyway, as the equation that determines the stationary values of $\br_T$ and $\br'_T$ is linear and independent of both $\br_o$ and the phase variation gradient, leading to the solution
\begin{equation}
\bar{\br}_T = \bar{\br}'_T = \frac{1}{z_a-z_b} \left[ (z_T+z_b-z) \frac{\br_a}{M_a} - (z_T+z_a-z) \frac{\br_b}{M_b} \right] .
\end{equation}
This enables to determine the stationary value of $\br_o$ through the remaining independent condition, leading to
\begin{equation}
\bar{\br}_o = - \frac{1}{z_a-z_b} \left[ (z-z_b) \frac{\br_a}{M_a} + (z_a-z) \frac{\br_b}{M_b}  \right] + z_T \bm{\nabla}\phi(\bar{\br}_T) .
\end{equation}
Therefore, despite the turbulence, the dominant contribution to the correlation function comes from a single object point,
\begin{multline}
\Gamma (\br_a,\br_b) \sim \mathcal{A}^2 \Biggl(  - \frac{1}{z_a-z_b} \left[ (z-z_b) \frac{\br_a}{M_a} + (z_a-z) \frac{\br_b}{M_b}  \right] \\ + z_T \bm{\nabla}\phi \left( \frac{1}{z_a-z_b} \left[ (z_T+z_b-z) \frac{\br_a}{M_a} - (z_T+z_a-z) \frac{\br_b}{M_b} \right] \right)  \Biggr) ,
\end{multline}
providing a \textit{single} shifted image for each pair of points $(\br_a,\br_b)$. Clearly, this image is detectable only provided the time scale of turbulence is slow enough to permit the reconstruction of the correlation function with a reasonable signal-to-noise ratio. Specializing the result to the case of reference, in which the object is focused on $\mathrm{D}_a$ (for definiteness), with $z=z_a$, the correlation function takes the interesting form
\begin{equation}\label{eq:Gammageo1}
\Gamma (\br_a,\br_b) \sim \mathcal{A}^2 \left(  - \frac{\br_a}{M_a} + z_T \bm{\nabla}\phi \left( \frac{1}{z_a-z_b} \left[ (z_T+z_b-z_a) \frac{\br_a}{M_a} - z_T \frac{\br_b}{M_b} \right] \right)  \right) ,
\end{equation}
the parallel with Eq.~\eqref{firstorder_geo} becomes clear: the choice of $(\br_a,\br_b)$ addresses a particular point on the turbulence plane, and therefore a specific shift of the image. Notice that the dependence of $\Gamma$ on $\br_b$ can be used as an indicator to detect the presence of turbulence with a transverse gradient. The situation when the turbulence plane is focused on $\mathrm{D}_b$, namely $z_b=z_a-z_T$,
\begin{equation}\label{eq:Gammageo2}
\Gamma (\br_a,\br_b) \sim \mathcal{A}^2 \left(  - \frac{\br_a}{M_a} + z_T \bm{\nabla}\phi \left( - z_T \frac{\br_b}{M_b} \right)  \right) ,
\end{equation}
becomes even simpler (and intuitive). Once the different sub-images, obtained by evaluating with varying $\br_b$, are collected, one can evaluate their relative allignment (e.g. by computing the point-to-point correlation between them) and follow two strategies to obtain an integrated image with higher signal-to-noise ratio:
\begin{itemize}
    \item sum only those sub-images characterized by the dominant allignment;
    \item realign all the sub-images and sum over them.
\end{itemize}

It is worth remarking that, due to the plenoptic properties of the correlation function, the advantage is not limited to the case of a focused object. However, though one can in principle reconstruct images of objects placed out of focus, wave-optics computation shows that the resolution of the correlation images is maximal in focus \cite{cpi_exp}. Here, resolution is determined by the Rayleigh limit, that is unattainable by traditional light-field techniques \cite{cpi_exp,massaro2022effect,massaro2022refocusing,scattarella2022resolution,scattarella2023periodic}.

\section{Discussion and outlook}

The results outlined in the previous section show that the light-field capability of CPI can be used to trace back the image of an object through phase disturbance, by using intensity correlations to isolate the contribution of a limited area of the turbulence plane. Compared to analogous applications of standard light-field imaging to turbulence mitigation \cite{wu2015imaging,wu2016imaging}, CPI potentially provides at the same time diffraction-limited resolution on the focused plane, and a much wider variety of independent viewpoints on a 3D sample \cite{cpi_prl}. Even though our analysis is limited to a focused object and to the geometrical-optics regime, recent wave-optics results demonstrated that the out-of-focus imaging properties of CPI entail a better volumetric resolution than any non-scanning first-order method \cite{massaro2022lightfield}.

One of the possible limitations of the method is related to the intrinsic need of CPI for collecting a large number of frames, to provide a stable statistical reconstruction of the correlation function. If the phase disturbance is not constant, the acquisition time must not exceed the typical variation time of the perturbation, to prevent different phase gradients from contributing to the same correlation evaluation [see Eqs.~\eqref{eq:Gammageo1}-\eqref{eq:Gammageo2}]. Keeping the acquisition time short, in turn, can lead to noisy images. Therefore, the technique can largely benefit from new sensors that combine high-resolution with low noise and gating times close to the nanosecond \cite{abbattista2021towards,antolovic2016photon,lubin2019quantum,ulku2020spad}, already applied to a CPI-AP experiment in a non-turbulent environment \cite{massaro2023correlated}. Moreover, further potential development can come from integrating the described methods with tools to maximize information extraction from data, like compressive sensing \cite{petrelli2023compressive} and artificial intelligence \cite{scattarella2023deep}, which can specifically help to realign noisy sub-images, even taken at different times.

Future research will be devoted to the analysis of phase disturbance effects on different methods of plenoptic imaging with intensity correlations, such as light-field ghost imaging \cite{paniate2024lightfield}), and to generalizing the results to different models of turbulence.

\end{document}